\documentclass[bimj,fleqn]{w-art}
\usepackage{times}
\usepackage{w-thm}
\usepackage[authoryear]{natbib}
\theoremstyle{plain}

\theoremstyle{definition}

\usepackage[]{graphicx}
\chardef\bslash=`\\ 

\begin{document}
\DOIsuffix{bimj.200100000}
\Volume{52}
\Issue{61}
\Year{2010}
\pagespan{1}{}
\keywords{Asymptotic theory; Genetics; High-dimensional data; Sequential goodness-of-fit; Type I error\\
\noindent \hspace*{-4pc} 
}

\title[Number of significant effects control]{Controlling the number of significant effects in multiple testing}
\author[Jacobo de U\~na-\'Alvarez]{Jacobo de U\~na-\'Alvarez\footnote{Corresponding author: {\sf{e-mail: jacobo@uvigo.gal}}, Phone: +34-986812492, Fax: +34-986812401}\inst{,1}} 
\address[\inst{1}]{CINBIO, Universidade de Vigo, 36310 Vigo, Spain}
\Receiveddate{zzz} \Reviseddate{zzz} \Accepteddate{zzz} 

\begin{abstract}
In multiple testing several criteria to control for type I errors exist. The false discovery rate, which evaluates the expected proportion of false discoveries among the rejected null hypotheses, has become the standard approach in this setting. However, false discovery rate control may be too conservative when the effects are weak. In this paper we alternatively propose to control the number of significant effects, where 'significant' refers to a pre-specified threshold $\gamma$. This means that a $(1-\alpha)$-lower confidence bound $L$ for the number of non-true null hypothesis with p-values below $\gamma$ is provided. When one rejects the nulls corresponding to the $L$ smallest p-values, the probability that the number of false positives exceeds the number of false negatives among the significant effects is bounded by $\alpha$. Relative merits of the proposed criterion are discussed. Procedures to control for the number of significant effects in practice are introduced and investigated both theoretically and through simulations. Illustrative real data applications are given.
\end{abstract}

\maketitle                   






\section{Introduction}

In multiple testing, the goal is to simultaneously test for $m$ null hypotheses $H_{0i}$, $1\leq i\leq m$, versus the corresponding alternative hypotheses $H_{1i}$, $1\leq i\leq m$. It is assumed that a set of independent and identically distributed p-values $p_i$, $1\leq i\leq m$, attached to the $m$ nulls is available, so a small $p_i$ indicate evidence against $H_{0i}$. Usually the researcher focuses on the $p_i$'s below a certain significance threshold $\gamma$. For example, in many instances he/she chooses $\gamma=0.05$. Then, the difficult task is to decide how many nulls to reject among those for which $p_i\leq \gamma$. This is because the multiple comparisons result in an inflated type I error rate. That is, while the critical region $\{p_i\leq \gamma\}$ individually controls the type I error probability at level $\gamma$, the probability of commiting one or more than one type I error along the rejected nulls may be overly large.\\

Most of the literature on type I error control in multiple testing has been focused on the familywise error rate (FWER) and false discovery rate (FDR) criteria; see for instance Benjamini and Hochberg (1995), Nichols and Hayasaka (2003) or Dudoit and Laan (2008). The FWER is defined as the probability of making at least one type I error along the $m$ tests. On its turn, the FDR is the expected proportion of false discoveries (these are, true nulls rejected) among the rejected null hypotheses. FWER-controlling procedures are often conservative. The statistical power is enhanced when controlling for the FDR, which is less restrictive. However, it has been pointed out that even FDR-controlling methods may be unable to provide a single discovery in particular situations like, for instance, the setting in which the effects are weak. In this work we follow a different line, looking for a better compromise between type I error control and power in such difficult scenarios.\\

To be specific, our focus will be on controlling at level $\alpha$ the probability that the number of rejected nulls is larger than the number of significant effects (NSE, in short), where we call \textit{significant effect} to a non-true null with p-value below the threshold $\gamma$. This is equivalent to protect oneself against the undesirable event that the amount of false positives exceeds the amount of false negatives with $p_i\leq \gamma$; see Section 2 for details. Clearly, the stated concept of NSE control is $\gamma$-dependent; we use the term \textit{$\gamma$-significant effect} to emphasized this fact. By focusing on $\gamma$-significant p-values one feels comfortable in the sense that, unlike when performing FDR control, null hypotheses with large p-values cannot be rejected. On the other hand, NSE control may greatly enhance the statistical power compared to FDR-controlling methods; evidence on this will be provided.\\


The rest of the paper is organized as follows. In Section 2 the new significance criterion is formally introduced. Methods to control for the NSE are proposed, and the main results are stated. In Section 3 a simulation study is performed in order to investigate and compare the practical behaviour of the introduced procedures. Section 4 is devoted to real data illustrations. A final discussion is given in Section 5. Proofs are collected in the Appendix.\\

\section{Methods and results}

\subsection{NSE control}

The new significance criterion requires a lower bound for the amount of effects with p-value below $\gamma$. Actually, a one-sided confidence interval with $1-\alpha$ confidence level is needed. To be precise, we look for an integer random variable $N^\alpha(\gamma)$ such that

\begin{equation}
	P(N^\alpha(\gamma)\leq mF_m^1(\gamma))\geq 1-\alpha
	\label{eq:key}
\end{equation}

\noindent under any configuration of the true and false null hypotheses, where $F_m^1(\gamma)=m^{-1}\sum_{i=1}^m I(p_i\leq \gamma, H_{0i}=1)$ denotes the proportion of p-values below $\gamma$ which are attached to non-true nulls ($H_{0i}=1$). Note that $mF_m^1(\gamma)$ is the number of effects (i.e. non-true nulls) among the nulls with p-value below the significance threshold $\gamma$ (that is, among the $\gamma$-significant p-values). Obviously, the random variable $N^\alpha(\gamma)$ must be supported on the set $\{0,\ldots,mF_m(\gamma)\}$, where $F_m(\gamma)=m^{-1}\sum_{i=1}^m I(p_i\leq \gamma)$. \\

\textbf{Definition 1 (NSE control).} \textit{A procedure that rejects the $N$ nulls with the smallest p-values is said to control the number of $\gamma$-significant effects (NSE) at level $\alpha$ if (a) $N\leq mF_m(\gamma)$ with probability one, and (b) (\ref{eq:key}) is satisfied with $N^\alpha(\gamma)=N$.}\\

Note that NSE control is defined for a specific, given significance threshold $\gamma$, so one should rather talk about $\gamma$-NSE control. However, in the following the reference to $\gamma$ will be often omitted for the sake of simplicity.\\

Equation (\ref{eq:key}) has a interesting connection with the weak control of the FWER and the FDR at level $\alpha$. Assume for a moment that all the nulls are true (complete null hypothesis). Then, in particular $mF_m^1(\gamma)=0$, and hence (\ref{eq:key}) can be written as $P(N^\alpha(\gamma)>0)\leq \alpha$. This means that any controlling procedure for the NSE ensures FWER control in the weak sense. The same holds true for the FDR, since FWER and FDR are equal under the complete null hypothesis. However, when some of the null hypotheses are false, equation (\ref{eq:key}) does not imply, nor is implied by, FWER (nor FDR) control at level $\alpha$.\\

An appealing feature of NSE is that it is equivalent to the strong control at level $\alpha$ for the undesirable event that the number of false positives exceeds the number of false negatives with $p_i\leq \gamma$. In order to see this, introduce the rejection indicator $R_i$, so $R_i=1$ when $H_{0i}$ is rejected, and $R_i=0$ otherwise, $1\leq i\leq m$. Then, the number of false positives is $\mbox{FP}=\sum_{i=1}^m I(R_i=1,H_{0i}=0)$, while the number of false negatives with $p_i\leq \gamma$ is $\mbox{FN}=\sum_{i=1}^m I(R_i=0,H_{0i}=1,p_i\leq \gamma)$. Here, $H_{0i}=0$ indicates that $H_{0i}$ is true. Now,

$$N^\alpha(\gamma)-mF_m^1(\gamma)=\sum_{i=1}^m I(R_i=1)-\sum_{i=1}^m I(p_i\leq \gamma,H_{0i}=1)$$

$$=\sum_{i=1}^m I(R_i=1,H_{0i}=0)+\sum_{i=1}^m I(R_i=1,H_{0i}=1)-\sum_{i=1}^m I(p_i\leq \gamma,H_{0i}=1)$$

$$=\mbox{FP }-\mbox{ FN},$$

\noindent where we have used that $R_i=1$ implies $p_i\leq \gamma$. Therefore, under (\ref{eq:key}) it holds $P(\mbox{FP}\leq \mbox{FN})\geq 1-\alpha$ or, equivalently, $P(\mbox{FP}>\mbox{FN})\leq \alpha$, as announced.\\

The aforementioned properties indicate in particular that, while FWER or FDR control may be violated by NSE-controlling procedures, the amount of rejected nulls under NSE control will not be unjustifiably large. In the following sections we introduce several procedures for NSE control.\\

Condition $P(\mbox{FP}>\mbox{FN})\leq \alpha$ imposed under NSE-control may resemble the generalized FWER criterion that bounds the probability that the number of false positives is greater than a given integer $k$. See for instance Dudoit and van der Laan (2008) for seminal methods controlling this so-called $k$-FWER criterion. However, in NSE-control the value of $k$ in $P(\mbox{FP}>k)\leq \alpha$ is replaced by the number of false negatives, FN, which is a random quantity. This entails an important difference with respect to $k$-FWER control, in which a preliminary, non-random choice of $k$ is performed. A related criterion is the false probability exceedance, which controls the tail probability of the false discovery proportion (FDP) at level $\alpha$. Here, the FDP is defined as FDP $=I(\mbox{R}>0)$ FP$/$R, where R stands for the number of rejected nulls; see D\"ohler and Roquain (2020) and references therein. Under such criterion, given a probability threshold $\xi \in (0,1)$, the inequality $P(\mbox{FDP}>\xi)\leq \alpha$ must be respected. Since, as the number of hypotheses $m$ grows, R $>0$ is eventually an almost sure event, this is asymptotically equivalent to imposing $P(\mbox{FP}>\xi \mbox{R})\leq \alpha$. Hence, control of the false probability exceedance has a similarity with NSE-control, but it uses a different (random) lower bound on the number of false positives. Despite of the aforementioned connections, NSE-cotrol works differently in that it looks for significant effects below a pre-specified p-value $\gamma$; no such threshold is imposed by $k$-FWER or FDP tail probability control.

\subsection{NSE controlling procedures: basic method}


Introduce $F^0_m(\gamma)=m^{-1}\sum_{i=1}^m I(p_i\leq \gamma,H_{0i}=0)$, so it holds $F_m(\gamma)=F_m^0(\gamma)+F_m^1(\gamma)$. Let $F^0(\gamma)=P(p_1\leq \gamma, H_{01}=0)$ be the expected value of $F_m^0(\gamma)$. Assume that the p-values corresponding to the true nulls are uniformly distributed; then, $F^0(\gamma)=P(p_1\leq \gamma|H_{01}=0)P(H_{01}=0)=\gamma \pi_0$, where $\pi_0$ stands for the a priori probability for a null hypothesis to be true. When $m$ is moderate to large, the Central Limit Theorem (CLT) provides an approximated $1-\alpha$ lower bound for $\gamma\pi_0$, given by

\begin{eqnarray*}
	F_m^0(\gamma)-z_\alpha \sqrt{\frac{\gamma \pi_0 (1 - \gamma \pi_0)}{m}},
\end{eqnarray*} 

\noindent where $z_\alpha$ is the $(1-\alpha)$-quantile of the standard normal. Since $F_m^1(\gamma)=F_m(\gamma)-F_m^0(\gamma)$, it holds

\begin{equation}
	P\left(F_m^1(\gamma)\geq F_m(\gamma)-\gamma \pi_0 - z_\alpha \sqrt{\frac{\gamma \pi_0 (1 - \gamma \pi_0)}{m}}\right)\rightarrow 1-\alpha,
	\label{eq:key2}
\end{equation}

\noindent as $m\rightarrow \infty$, from which the following result is obtained.\\

\textbf{Proposition 1.} \textit{The rule $N^\alpha(\gamma;\pi_0)=mF_m(\gamma)-m\gamma\pi_0 - z_\alpha (m\gamma \pi_0(1-\gamma \pi_0))^{1/2}$ controls the NSE at level $\alpha$ as $m \rightarrow \infty$.}\\

Actually, we propose to consider the integer part of $N^\alpha(\gamma;\pi_0)$ in Proposition 1 in order to make a decision (same applies to other methods to be introduced in the following). On the other hand, note that $N^\alpha(\gamma;\pi_0) \leq mF_m(\gamma)$ and, therefore, $N^\alpha(\gamma;\pi_0)$ satisfies the support condition required by Definition 1.\\

The rule in Proposition 1 may be redefined in a obvious way to ensure NSE control for a fixed value of $m$, and not only asymptotically. For this, one must use the $(1-\alpha)$-quantile of a binomial, $B(m,\gamma\pi_0)$ distribution, $b_{m,\alpha}(\gamma \pi_0)$ say, rather than of a standard normal. In fact, the asymptotic result in Proposition 1 can be derived from the usual binomial-normal approximation

$$b_{m,\alpha}(\gamma\pi_0) \simeq m\gamma \pi_0 + z_\alpha (m\gamma \pi_0(1-\gamma\pi_0))^{1/2},$$

\noindent which holds for large $m$. Results in the rest of the paper will focus on the asymptotic setting in which $m \rightarrow \infty$.\\

The rule $N^\alpha(\gamma;\pi_0)$ is unpractical since it involves the unknown parameter $\pi_0$. At this point two different strategies are possible. The first one is the conservative decision that is obtained by setting $\pi_0=1$. When $\gamma \leq 1/2$ (this will be the case in most applications), it holds $\gamma\pi_0(1-\gamma \pi_0)\leq \gamma(1-\gamma)$. Then, from (\ref{eq:key2}) we obtain

\begin{equation*}
	P(N^\alpha(\gamma;1)\leq mF_m^1(\gamma)) \rightarrow \nu
\end{equation*}

\noindent as $m \rightarrow \infty$, where $\nu\geq 1-\alpha$. Indeed, it is easily seen that, when $\pi_0<1$, it holds $\nu= 1$. This indicates that, when there exists a positive fraction of non-true nulls, $N^\alpha(\gamma;1)$ asymptotically provides an almost sure lower bound for the number of effects with p-value below the threshold $\gamma$.\\

Rule $N^\alpha(\gamma;1)=mF_m(\gamma)-m\gamma - z_\alpha (m\gamma (1-\gamma))^{1/2}$ equals the asymptotic version of SGoF procedure as defined in de U\~na-\'Alvarez (2011); see Carvajal-Rodr\'iguez et al. (2009) for the original conception of SGoF as a sequential goodness-of-fit method. Therefore, SGoF is a NSE-controlling method. The same applies to the conservative version of SGoF that replaces the term $\gamma(1-\gamma)$ under the square root in $N^\alpha(\gamma;1)$ by $F_m(\gamma)(1-F_m(\gamma))$; this is formally true at least when $\gamma \leq F_m(\gamma) \leq 1/2$, which is a realistic assumption (de U\~na-\'Alvarez, 2012; Castro-Conde and de U\~na-\'Alvarez, 2015). SGoF-type methods are related to the notion of second-level significance testing, or higher criticism, introduced by Tukey in 1976, and exploited later by Donoho and Jin (2004), see also Hall and Jin (2008). The basic idea behind SGoF is to compare the number of $\gamma$-significant p-values to the expected amount under the complete null. SGoF procedures have become popular in applied research due to their ability to detect non-true nulls in settings with a small proportion of weak effects, when FDR control may be too restrictive (Diz et al. 2011, Mart\'inez-Camblor 2014, Polewko-Klim et al. 2021, Ni et al. 2023); their friendly-use implementation in the \texttt{sgof} package (Castro-Conde and de U\~na-\'Alvarez, 2014) has boosted the dissemination too.\\


The second strategy to introduce a practical version of $N^\alpha(\gamma;\pi_0)$ is to replace the unknown $\pi_0$ by a suitable estimator $\hat \pi_0$. This leads to a decision rule $N^\alpha(\gamma;\hat \pi_0)$ for which $P(N^\alpha(\gamma;\hat \pi_0)\leq mF_m^1(\gamma))$ is expected to be closer to the nominal $1-\alpha$ than when working with $\pi_0=1$; simulations below confirm this guess. In order to guarantee $P(N^\alpha(\gamma;\hat \pi_0)\leq mF_m^1(\gamma)) \geq 1-\alpha$ it is important to plug-in a conservative estimator $\hat \pi_0$, so it tends to take values larger than the true $\pi_0$. This is the case for two simple estimators of $\pi_0$ (among other existing estimators): the one proposed by Storey (2002), and the one introduced by Dalmasso et al. (2005). We review these two approaches in the following subsection.\\

Interestingly, compared to SGoF, NSE control with plug-in rules may provide a quite larger statistical power in practice. This is because plug-in NSE-controlling procedures estimate the expected number of $\gamma$-significant p-values under any configuration of the null hypotheses, and not only under the complete null. Another advantage is that NSE focuses on the \textit{actual} number of significant effects, being liberal with respect to the control of the \textit{expected} amount, which is the target of SGoF methods. This leads to some variance improvements and, consequently, to power enhancements. See Section 3 for discussion in particular simulation scenarios, and Section 4 for illustrative comparisons with real data.\\

\subsection{NSE controlling procedures: plug-in rules}

In this subsection we introduce plug-in rules for NSE control. Storey (2002) proposed to estimate $\pi_0$ by

\begin{equation}
	\hat \pi_0^S=\frac{1}{m(1-\lambda)}\sum_{i=1}^m I(p_i>\lambda)=\frac{1-F_m(\lambda)}{1-\lambda}
	\label{eq:storey}
\end{equation}

\noindent for some well-chosen $\lambda$. The rationale for (\ref{eq:storey}) is that most of the p-values larger than $\lambda$ (when $\lambda$ is properly chosen) should correspond to true nulls and that, on the other hand, null p-values are uniformly distributed on the unit interval. Put $F$ for the common cumulative distribution function (cdf) of the p-values. The expected value of $\hat \pi_0^S$ is

\begin{equation*}
	E(\hat \pi_0^S)=\frac{1-F(\lambda)}{1-\lambda}=\pi_0+\frac{(1-\pi_0)(1-F^{(1)}(\lambda))}{1-\lambda},
\end{equation*}

\noindent where $F^{(1)}(\cdot)=P(p_1\leq \cdot|H_{01}=1)$ denotes the cdf of the p-values attached to the non-true nulls, and where we have used $F(x)=\pi_0x+(1-\pi_0)F^{(1)}(x)$, $0\leq x\leq 1$. This shows that (\ref{eq:storey}) tends to overestimate the target. Storey (2002) suggested to use $\lambda=1/2$ as a simple rule to select the value of $\lambda$; this is the version of Storey's method considered in the current paper.\\

An alternative estimator of $\pi_0$ is given by

\begin{equation}
	\hat \pi_0^D=-\frac{1}{m}\sum_{i=1}^m \log(1-p_i) =-\int \log(1-x)dF_m(x).
	\label{eq:dalmasso}
\end{equation}

\noindent The estimator (\ref{eq:dalmasso}) was proposed by Dalmasso et al. (2005) within a general class of estimators. It holds

\begin{equation*}
	E(\hat \pi_0^D)=\pi_0-(1-\pi_0)\int \log(1-x)dF^{(1)}(x)
\end{equation*}

\noindent which reveals that, similarly to $\hat \pi_0^S$, $\hat \pi_0^D$ is a conservative estimator of $\pi_0$.\\

The rules $N^\alpha(\gamma;\hat \pi_0^S)$ and $N^\alpha(\gamma;\hat \pi_0^D)$ are obtained from the NSE-controlling procedure $N^\alpha(\gamma;\pi_0)$ by estimating the parameter $\pi_0$. In principle, the fact that $\pi_0$ is being estimated could lead to a violation of NSE control; simulations in Section 3 confirm this. In order to avoid the issue, the randomness of $\hat \pi_0^S$ and $\hat \pi_0^D$ should be taken into account. This results in two new methods, with updated variance terms, for which an asymptotic NSE control can be theoretically established. The following two Propositions formalize things. The proofs are deferred to the Appendix.\\

\textbf{Proposition 2.} \textit{The rule $N_S^\alpha(\gamma)=mF_m(\gamma)-m\gamma\hat \pi_0^S - z_\alpha (m \sigma^2_S(\gamma;\pi_0))^{1/2}$, where}

\begin{eqnarray*}
	\sigma^2_S(\gamma;\pi_0)=\gamma \pi_0+\frac{\gamma^2}{(1-\lambda)^2}(1-F(\lambda))-\frac{2\gamma}{1-\lambda}I(\lambda<\gamma)(\gamma-\lambda)\pi_0\\
	-\left(\gamma\pi_0-\frac{\gamma(1-F(\lambda))}{1-\lambda}\right)^2,
\end{eqnarray*}

\noindent \textit{controls the NSE at level $\alpha$ as $m \rightarrow \infty$.}\\

\textbf{Proposition 3.} \textit{The rule $N_D^\alpha(\gamma)=mF_m(\gamma)-m\gamma\hat \pi_0^D - z_\alpha (m \sigma^2_D(\gamma;\pi_0))^{1/2}$, where}

\begin{eqnarray*}
	\sigma^2_D(\gamma;\pi_0)=\gamma\pi_0+\gamma^2 \int \log^2(1-x)dF(x)-2\gamma\pi_0[(1-\gamma)\log(1-\gamma)+\gamma]\\
	-\left( \gamma\pi_0+\gamma\int \log(1-x)dF(x)\right)^2,
\end{eqnarray*}

\noindent \textit{controls the NSE at level $\alpha$ as $m \rightarrow \infty$.}\\

The variances $\sigma^2_S(\gamma;\pi_0)$ and $\sigma^2_D(\gamma;\pi_0)$ in Propositions 2 and 3 are generally larger than the variance in Proposition 1, namely $\sigma^2(\gamma;\pi_0)= \gamma\pi_0(1- \gamma\pi_0)$. This variance increase that arises from the estimation of $\pi_0$ may be particularly visible for moderate to large values of $\gamma$, or when there exists a small proportion of relatively weak effects, that is, true alternative hypotheses that are close to their corresponding non-true nulls. Therefore, the updated variances introduced in Propositions 2 and 3 are critical for NSE control; see Section 3 for results and further discussion. In practice, $\sigma^2_S(\gamma;\pi_0)$ and $\sigma^2_D(\gamma;\pi_0)$ can be easily estimated by replacing $\pi_0$ and $F$ by $\hat \pi_0^S$ (respectively $\hat \pi_0^D$) and $F_m$. Other estimators for $\pi_0$ could be used for this aim too.\\

Specifically, the variance estimators considered in the following section are as follows:

$$\hat \sigma^2_S(\gamma)=\gamma \hat \pi_0^S+\frac{\gamma^2}{1-\lambda}\hat \pi_0^S-\frac{2\gamma}{1-\lambda}I(\lambda<\gamma)(\gamma-\lambda)\hat \pi_0^S$$

\noindent for $\sigma^2_S(\gamma;\pi_0)$, and

$$\hat \sigma^2_D(\gamma)=\gamma\hat \pi_0^D+\gamma^2 \int \log^2(1-x)dF_m(x)-2\gamma\hat \pi_0^D[(1-\gamma)\log(1-\gamma)+\gamma]$$

\noindent for $\sigma^2_D(\gamma;\pi_0)$. Note that the last term in $\sigma^2_S(\gamma;\pi_0)$ (Proposition 2), respectively in $\sigma^2_D(\gamma;\pi_0)$ (Proposition 3), vanishes when plugging-in the corresponding estimators; this is why the term is not appearing in the expressions of $\hat \sigma^2_S(\gamma)$ and $\hat \sigma^2_D(\gamma)$. On the other hand, when $\lambda=1/2$ the estimator $\hat \sigma^2_S(\gamma)$ reduces to $\hat \sigma^2_S(\gamma)=\gamma \hat \pi_0^S(1+\gamma/(1-\lambda))$ whenever $\gamma\leq 1/2$, which will be often the case in applications.

\section{Simulation study}

In this section we investigate the performance of the proposed NSE-controlling procedures through simulations. The methods we consider are the following:

\begin{itemize}
	\item [(a)] The benchmark method based on the true (unavailable in practice) value of $\pi_0$, that is, $N^\alpha(\gamma;\pi_0)$;
	\item [(b)] The conservative approach which sets $\pi_0=1$, that is, $N^\alpha(\gamma;1)$, which coincides with original SGoF (Carvajal-Rodr\'guez et al., 2009);
	\item [(c)] The plug-in approach based on Storey (2002)'s estimator for $\pi_0$ ($\lambda=1/2$), with no update for the variance term; this is $N^\alpha(\gamma;\hat \pi_0^S)$;
	\item [(d)] The plug-in approach based on Dalmasso et al. (2005)'s estimator for $\pi_0$, with no update for the variance term; this is $N^\alpha(\gamma;\hat \pi_0^D)$;
	\item [(e)] The plug-in approach based on Storey (2002)'s estimator for $\pi_0$ ($\lambda=1/2$), with updated variance term; this is $N_S^\alpha(\gamma)$, which replaces the term $\sigma^2(\gamma;\hat \pi_0^S)= \gamma\hat \pi_0^S(1-\hat \gamma\pi_0^S)$ in $N^\alpha(\gamma;\hat \pi_0^S)$ by $\hat \sigma^2_S(\gamma)$;
	\item [(f)] The plug-in approach based on Dalmasso et al. (2005)'s estimator for $\pi_0$, with updated variance term; this is $N_D^\alpha(\gamma)$, which replaces the term $\sigma^2(\gamma;\hat \pi_0^D)= \gamma\hat \pi_0^D(1-\gamma\hat \pi_0^D)$ in $N^\alpha(\gamma;\hat \pi_0^D)$ by $\hat \sigma^2_D(\gamma)$.
\end{itemize}

The simulated scenario corresponds to a two-sided test for the mean based on a random sample with $n=5$ observations drawn from a Gaussian population with known variance, specifically a $N(\mu,1)$ model. Therefore, the p-value can be simulated as $p_i=2\Phi(-|Z_i|)$, where $\Phi$ denotes the cdf of the standard normal and $Z_i$ follows a $N(\sqrt{n}\mu,1)$ model. The null hypothesis is $H_{0i}:\mu=0$ and the alternative hypothesis is $H_{1i}:\mu \neq 0$. We simulate $m$ testing problems ($m \in \{1000,10000\}$) with a random proportion of true nulls with average $\pi_0$. Chosen values for $\pi_0$ are $1$ (complete null), $0.9$ (10\% of non-true nulls) and $0.8$ (20\% of non-true nulls). For the non-true nulls the value of $\mu$ is set to $1$, $1.5$ or $2$ to represent weak, moderate or strong departures from the null (individual power: $0.6088$, $0.9184$ and $0.9940$). The significance thresholds are $\gamma \in \{0.05,0.01\}$, and the NSE is controlled at levels $\alpha \in \{0.05,0.01\}$. The number of Monte Carlo replicates is $5000$.\\

In Tables 1, 3, 4 and 5 we report the coverage for the NSE with several combinations $(m,\gamma,\alpha)$. This NSE coverage is defined as the proportion of trials for which the number of rejected nulls was not larger than the number of effects with p-value below $\gamma$; note that this proportion should be equal to, or larger than, $1-\alpha$. Closeness of such proportion to the nominal $1-\alpha$ is intended in any case. For completeness, we include in Tables 1, 3, 4 and 5 the FDR (computed as the average proportion of false discoveries along the Monte Carlo trials) and the power (computed as the average proportion of discovered effects). A small value for the FDR is appealing; however, the goal here is to maximize the power while controlling the NSE at level $\alpha$.\\

From Tables 1, 3, 4 and 5 it is seen that original SGoF rule $N^\alpha(\gamma;1)$ is very conservative whenever $\pi_0<1$. This was already anticipated in Section 2. On the other hand, methods $N^\alpha(\gamma;\hat \pi_0^S)$ and $N^\alpha(\gamma;\hat \pi_0^D)$, with no variance update for the estimation of $\pi_0$, fail to respect the nominal level. Rules $N_S^\alpha(\gamma)$ and $N_D^\alpha(\gamma)$ perform well, respecting the level and providing statistical power close to that of the benchmark method. Among these latter methods, a slightly better performance is found for the one based on Storey's estimator. This can be explained from the fact that, in the simulated scenarios, Dalmasso's estimator was more conservative than Storey's, particularly when the effects were weak ($\mu=1$); see Table \ref{tab:pi0_m1000}. Methods (e) and (f) were somehow conservative for weak effects and $\pi_0=0.8$, when they reported coverages above the nominal $1-\alpha$ (more evident for $m=10000$, Table 5); this is because of the relative size of $-\mu_T/\sigma_T$, see Table \ref{tab:musigma} in the Appendix, which induces a departure from the $(1-\alpha)$-normal quantile $z_\alpha$. Note that the value of $\mu_T$ is zero under the complete null ($\pi_0=1$), when $\hat \pi_0^S$ and $\hat \pi_0^D$ are perfectly unbiased, but not in general.\\

It is also clear from Tables 1, 3, 4 and 5 that the power increases with $\alpha$, the effect size $\mu$ and the proportion of non-true nulls $1-\pi_0$; this was expected. The impact of $\gamma$ in the power is less evident; for instance, the choice $\gamma=0.01$ reports a smaller power (compared to $\gamma=0.05$) when $\mu\leq 1.5$, but the opposite occurs when $\mu=2$. A bias-variance trade-off is responsible for this.\\

NSE control refers to bouding the \textit{actual} number of $\gamma$-significant effects. None of the methods in the simulation study control, however, for the \textit{expected} number of effects with p-value below the $\gamma$-threshold. In order to see this, we computed the Monte Carlo average for the event $\{N \leq mF^1(\gamma)\}$, where $F^1(\gamma)=P(p_1\leq \gamma, H_{01}=1)$, for one of the simulated scenarios. Explicitly, for $\pi_0=0.9$, $\alpha=\gamma=0.05$, $\mu=2$ and $m=1000$, the referred Monte Carlo average was below $86\%$ for all the methods, with the only exception of $N^\alpha(\gamma;1)$, which reached $93\%$. See Remarks 3 and 4 in the Appendix for further discussion.\\

\begin{table}[htb]
	\begin{center}
			\caption{NSE coverage, FDR and power for methods (a)-(f) along 5000 Monte Carlo trials, depending on the proportion of true nulls ($\pi_0$) and the degree of violation of the null hypothesis $\mu=0$. Case $\gamma=\alpha=0.05$ and $m=1000$.}
		\begin{tabular}{ cccccccc } 
			\hline
			& 	 & (a) & (b) & (c) & (d) & (e) & (f) \\
			\hline
			$\pi_0=1$ & & & & & & & \\
			\hline
			& NSE & 0.9598 & 0.9598 & 0.9468 & 0.9460 & 0.9584 & 0.9570\\
			\hline
			$\pi_0=0.9$ & & & & & & & \\
			\hline
			$\mu=1$& NSE	 & 0.9546 & 0.9952 & 0.9558 & 0.9614 & 0.9644 & 0.9678\\ 
			& FDR	 & 0.2159 & 0.1883 & 0.2111 & 0.2102 & 0.2076 & 0.2066\\ 
			& pow	 & 0.3894 & 0.3540  & 0.3835 & 0.3822 & 0.3791 & 0.3776\\ 
			& & & & & & & \\
			$\mu=1.5$& NSE	 & 0.9546 & 0.9952 & 0.9494 & 0.9514 & 0.9588 & 0.9620\\ 
			& FDR	 & 0.0815 & 0.0617 & 0.0805 & 0.0804 & 0.0778 & 0.0777\\ 
			& pow	 & 0.7410 & 0.7006 & 0.7381 & 0.7376 & 0.7331 & 0.7326\\
			& & & & & & & \\
			$\mu = 2$& NSE	 & 0.9546 & 0.9952 & 0.9482 & 0.9494 & 0.9582 & 0.9616\\ 
			& FDR	 & 0.0149 & 0.0071 & 0.0152 & 0.0151 & 0.0138 & 0.0137\\ 
			& pow	 & 0.8706 & 0.8180 & 0.8669 & 0.8671 & 0.8605 & 0.8606\\  
			\hline
			$\pi_0=0.8$&  & & & & & & \\
			\hline
			$\mu=1$& NSE	 & 0.9656 & 0.9996 & 0.9650 & 0.9734 & 0.9714 & 0.9782\\ 
			& FDR	 & 0.1509 & 0.1307 & 0.1491 & 0.1480 & 0.1477 & 0.1467\\ 
			& pow	 & 0.4700 & 0.4333 & 0.4668 & 0.4650 & 0.4643 & 0.4627\\ 
			& & & & & & & \\
			$\mu=1.5$& NSE	 & 0.9656 & 0.9996 & 0.9496 & 0.9536 & 0.9596 & 0.9642\\ 
			& FDR	 & 0.0626 & 0.0461 & 0.0631 & 0.0628 & 0.0618 & 0.0616\\ 
			& pow	 & 0.8090 & 0.7708 & 0.8096 & 0.8090 & 0.8071 & 0.8067\\
			& & & & & & & \\
			$\mu=2$& NSE	 & 0.9656 & 0.9996 & 0.9480 & 0.9498 & 0.9588 & 0.9606\\ 
			& FDR	 & 0.0136 & 0.0054 & 0.0143 & 0.0142 & 0.0134 & 0.0134\\ 
			& pow	 & 0.9264 & 0.8794 & 0.9272 & 0.9271 & 0.9243 & 0.9243\\ 
			\hline
		\end{tabular}
	\end{center}
	\label{tab:simg05a05m1000}
\end{table}

\begin{center}
	\begin{table}
		\begin{center}
					\caption{Mean and standard deviation of the estimators $\hat \pi_0^S$ and $\hat \pi_0^D$ along 5000 Monte Carlo trials depending on the proportion of true nulls ($\pi_0$) and the degree of violation of the null hypothesis $\mu=0$. Case $m=1000$. Truncation at 1, whenever needed, was performed for admissibility reasons.}
			\begin{tabular}{ cccccc } 
				\hline
				$\pi_0$& $\mu$ &	  $E(\hat \pi_0^S)$ & sd$(\hat \pi_0^S)$ & $E(\hat \pi_0^D)$ & sd$(\hat \pi_0^D)$  \\
				\hline
				& & & & & \\
				$1$	 & $-$ & 0.9873 & 0.0184 & 0.9873 & 0.0181\\
				& & & & & \\
				$0.9$	& 	$1$	  & 0.9110 & 0.0315 & 0.9160 & 0.0311\\ 
				&	$1.5$	   & 0.9003 & 0.0314 & 0.9019 & 0.0312\\ 
				&	$2$	 & 0.8996  & 0.0314 & 0.8999 & 0.0312\\ 
				& & & & & \\
				$0.8$ &	$1$	   & 0.8226 & 0.0311 & 0.8322 & 0.0304\\ 
				&	$1.5$	   & 0.8011 & 0.0309 & 0.8040 & 0.0306\\ 
				&	$2$	   & 0.7997 & 0.0308 & 0.8000 & 0.0306\\ 
				& & & & & \\
				\hline
			\end{tabular}
		\end{center}
		\label{tab:pi0_m1000}
	\end{table}
\end{center}

\begin{table}
	\begin{center}
			\caption{NSE coverage, FDR and power for methods (a)-(f) along 5000 Monte Carlo trials, depending on the proportion of true nulls ($\pi_0$) and the degree of violation of the null hypothesis $\mu=0$. Case $\gamma=0.01$, $\alpha=0.05$ and $m=1000$.}
		\begin{tabular}{ cccccccc } 
			\hline
			& 	 & (a) & (b) & (c) & (d) & (e) & (f) \\
			\hline
			$\pi_0=1$ & & & & & & & \\
			\hline
			& NSE &  0.9746 & 0.9746 & 0.9662 & 0.9654 & 0.9678 & 0.9682\\
			\hline
			$\pi_0=0.9$ & & & & & & & \\
			\hline
			$\mu=1$& NSE	 & 0.9624 & 0.9910  & 0.9714 & 0.9740 & 0.9732 & 0.9750\\ 
			& FDR	 & 0.1299 & 0.1211 & 0.1273 & 0.1271 & 0.1269 & 0.1268\\ 
			& pow	 & 0.2747 & 0.2597 & 0.2704 & 0.2699 & 0.2698 & 0.2694\\
			& & & & & & & \\ 
			$\mu=1.5$& NSE	 & 0.9624 & 0.9910 & 0.9694 & 0.9696 & 0.9710 & 0.9718\\ 
			& FDR	 & 0.0539 & 0.0485  & 0.0528 & 0.0526 & 0.0525 & 0.0524\\ 
			& pow	 & 0.6906 & 0.6754 & 0.6875& 0.6873 & 0.6868 & 0.6867\\
			& & & & & & & \\
			$\mu=2$& NSE	 & 0.9624 & 0.9910 & 0.9694 & 0.9690 & 0.9708 & 0.9708\\ 
			& FDR	 & 0.0180 & 0.0139 & 0.0172 & 0.0172 & 0.0170 & 0.0170\\ 
			& pow	 & 0.9035 & 0.8876 & 0.9003 & 0.9003 & 0.8996 & 0.8996\\  
			\hline
			$\pi_0=0.8$&  & & & & & & \\
			\hline
			$\mu=1$& NSE	 & 0.9670 & 0.9976  & 0.9730 & 0.9748 & 0.9736 & 0.9764\\ 
			& FDR	 & 0.0768 & 0.0721 & 0.0761 & 0.0758 & 0.0759 & 0.0757\\ 
			& pow	 & 0.3153 & 0.3029 & 0.3136 & 0.3131 & 0.3133 & 0.3128\\ 
			& & & & & & & \\
			$\mu=1.5$& NSE	 & 0.9670 & 0.9976 & 0.9676 & 0.9674 & 0.9694 & 0.9682\\ 
			& FDR	 & 0.0330 & 0.0299 & 0.0328 & 0.0328 & 0.0327 & 0.0327\\ 
			& pow	 & 0.7308 & 0.7185 & 0.7302 & 0.7300 & 0.7299 & 0.7298\\
			& & & & & & & \\
			$\mu=2$& NSE	 & 0.9670 & 0.9976 & 0.9674 & 0.9666 & 0.9688 & 0.9676\\ 
			& FDR	 & 0.0134 & 0.0103 & 0.0133 & 0.0132 & 0.0132 & 0.0132\\ 
			& pow	 & 0.9331 & 0.9211 & 0.9326 & 0.9325 & 0.9323 & 0.9323\\ 
			\hline
		\end{tabular}
	\end{center}
	\label{tab:simg01a05m1000}
\end{table}

\begin{table}
	\begin{center}
			\caption{NSE coverage, FDR and power for methods (a)-(f) along 5000 Monte Carlo trials, depending on the proportion of true nulls ($\pi_0$) and the degree of violation of the null hypothesis $\mu=0$. Case $\gamma=0.05$, $\alpha=0.01$ and $m=1000$.}
		\begin{tabular}{ cccccccc } 
			\hline
			& 	 & (a) & (b) & (c) & (d) & (e) & (f) \\
			\hline
			$\pi_0=1$ & & & & & & & \\
			\hline
			& NSE & 0.9916 & 0.9916 & 0.9844 & 0.9850 & 0.9896 & 0.9890\\
			\hline
			$\pi_0=0.9$ & & & & & & & \\
			\hline
			$\mu=1$& NSE	 & 0.9928  & 0.9994 & 0.9906 & 0.9916 & 0.9936 & 0.9952\\ 
			& FDR	 & 0.1928 & 0.1658  & 0.1910 & 0.1897 & 0.1858 & 0.1845\\ 
			& pow	 & 0.3601 & 0.3218  & 0.3570 & 0.3552 & 0.3499 & 0.3483\\
			& & & & & & & \\ 
			$\mu=1.5$& NSE	 & 0.9928 & 0.9994 & 0.9882 & 0.9888 & 0.9920 & 0.9928\\ 
			& FDR	 & 0.0647 & 0.0481  & 0.0657 & 0.0656 & 0.0623 & 0.0621\\ 
			& pow	 & 0.7077 & 0.6631 & 0.7084 & 0.7077 & 0.7004 & 0.7000\\
			& & & & & & & \\
			$\mu=2$& NSE	 & 0.9928 & 0.9994 & 0.9880 & 0.9884 & 0.9920 & 0.9918\\ 
			& FDR	 & 0.0081 & 0.0038  & 0.0089 & 0.0087 & 0.0076 & 0.0077\\ 
			& pow	 & 0.8271 & 0.7707  & 0.8286 & 0.8285 & 0.8185 & 0.8185\\  
			\hline
			$\pi_0=0.8$&  & & & & & & \\
			\hline
			$\mu=1$& NSE	 & 0.9910 & 1.000 &0.9910 & 0.9932& 0.9936&0.9950\\ 
			& FDR	 & 0.1434 & 0.1217 & 0.1411& 0.1400&0.1392 &0.1382\\ 
			& pow	 & 0.4570 & 0.4157 & 0.4526&0.4508 &0.4492 &0.4473\\ 
			& & & & & & & \\
			$\mu=1.5$& NSE	 & 0.9910 & 1.000 &0.9864 &0.9872 & 0.9906&0.9908\\ 
			& FDR	 & 0.0561 & 0.0398 & 0.0561& 0.0559&0.0546 &0.0544\\ 
			& pow	 & 0.7957 & 0.7518 &0.7955 &0.7949 & 0.7920&0.7915\\
			& & & & & & & \\
			$\mu=2$& NSE	 & 0.9910 & 1.000 &0.9858 &0.9864 & 0.9906&0.9900\\ 
			& FDR	 & 0.0097 & 0.0037 &0.0100 &0.0100 & 0.0092&0.0092\\ 
			& pow	 & 0.9103 & 0.8560 & 0.9101& 0.9101& 0.9060&0.9059\\ 
			\hline
		\end{tabular}
	\end{center}
	\label{tab:simg05a01m1000}
\end{table}

\begin{table}
	\begin{center}
			\caption{NSE coverage, FDR and power for methods (a)-(f) along 5000 Monte Carlo trials, depending on the proportion of true nulls ($\pi_0$) and the degree of violation of the null hypothesis $\mu=0$. Case $\gamma=\alpha=0.05$ and $m=10000$.}
		\begin{tabular}{ cccccccc } 
			\hline
			& 	 & (a) & (b) & (c) & (d) & (e) & (f) \\
			\hline
			$\pi_0=1$ & & & & & & & \\
			\hline
			& NSE & 0.9524&0.9524 &0.9366 & 0.9360& 0.9526&0.9528 \\
			\hline
			$\pi_0=0.9$ & & & & & & & \\
			\hline
			$\mu=1$& NSE	 & 0.9562 &1.000  &0.9666 & 0.9740& 0.9736&0.9772\\ 
			& FDR	 & 0.2495 & 0.2270 & 0.2471&0.2460 &0.2460 &0.2449\\ 
			& pow	 &0.4300  & 0.4034 &0.4271 & 0.4259& 0.4259&0.4246\\
			& & & & & & & \\ 
			$\mu=1.5$& NSE	 & 0.9562 & 1.000  & 0.9454 & 0.9492 & 0.9558 & 0.9598\\ 
			& FDR	 & 0.1084 & 0.0876 & 0.1084 & 0.1081 & 0.1073 & 0.1070\\ 
			& pow	 & 0.7873 & 0.7591  & 0.7872 & 0.7868 & 0.7859 & 0.7855\\
			& & & & & & & \\
			$\mu=2$& NSE	 & 0.9562 & 1.000 & 0.9432&0.9448 & 0.9542&0.9566\\ 
			& FDR	 &0.0284  & 0.0138 & 0.0287& 0.0287&0.0278 &0.0277\\ 
			& pow	 & 0.9313 & 0.8951 & 0.9314& 0.9313&0.9299 &0.9298\\  
			\hline
			$\pi_0=0.8$&  & & & & & & \\
			\hline
			$\mu=1$& NSE	 & 0.9576 & 1.000 & 0.9806& 0.9894&0.9850 &0.9912\\ 
			& FDR	 & 0.1650 & 0.1458 & 0.1629&0.1619 & 0.1624&0.1615\\ 
			& pow	 & 0.4942 & 0.4615 & 0.4906&0.4890 & 0.4899&0.4883\\ 
			& & & & & & & \\
			$\mu=1.5$& NSE	 & 0.9576 & 1.000 &0.9500 &0.9576 &0.9612 &0.9654\\ 
			& FDR	 & 0.0758 & 0.0575 &0.0757 & 0.0754& 0.0753&0.0750\\ 
			& pow	 & 0.8333 & 0.8013 &0.8332 & 0.8327& 0.8325&0.8321\\
			& & & & & & & \\
			$\mu=2$& NSE	 & 0.9576 & 1.000 & 0.9464& 0.9446&0.9576 &0.9582\\ 
			& FDR	 &0.0230  & 0.0091 & 0.0231&0.0230 &0.0226 &0.0226\\ 
			& pow	 & 0.9548 & 0.9174 & 0.9548&0.9548 & 0.9542&0.9541\\ 
			\hline
		\end{tabular}
	\end{center}
	\label{tab:simg05a05m10000}
\end{table}

The FDR ranged between 1\% and 25\% in the performed simulations, depending on the effect size, the proportion of true null hypotheses, the number of tests, and the particular choices for $\gamma$ and $\alpha$. This means that the application of a FDR-controlling procedure at the usual levels would have provided a larger or smaller statistical power compared to NSE-controlling methods depending on the scenario. This is an interesting aspect of NSE control, since it automatically adapts to the relative difficulty when looking for effects. For example, FDR control at nominal level 5\% would have performed poorly in our scenarios of weak effects ($\mu=1$) with a small proportion of non-true nulls ($1-\pi_0=0.1$), where the average proportion of false discoveries for NSE-controlling procedures was always above 12\%. On the other hand, the existence of two significance parameters $\gamma$ and $\alpha$ in NSE control provides the researcher with an extra flexibility compared to FDR-based methods.

\section{Real data illustrations}

\subsection{Hedenfalk breast cancer data}

Hedenfalk et al. (2001) performed a study of hereditary breast cancer by considering, among other things, gene expression levels measured from patients with two different types of mutations: BRCA1 (7 cases) and BRCA2 (8 cases). We focus on the $m=3170$ p-values arising from the comparison of the two groups, see Storey and Tibshirani (2003) for further details. The data are available from \texttt{Hedenfalk} within the package \texttt{sgof}.\\


For Hedenfalk data, the estimated proportion of non-true nulls is about 30\% ($\hat \pi_0^S=0.6763$, $\hat \pi_0^D=0.7177$). We applied the proposed NSE controlling procedures with initial significance thresholds $\gamma=0.05$ and $\gamma=0.01$. The number of significant p-values was $606$ ($\gamma=0.05$) or $265$ ($\gamma=0.01$). The level $\alpha$ was set to $0.05$ or $0.01$. The results are displayed in Table \ref{tab:hedenfalk}. From Table \ref{tab:hedenfalk} it is seen that the role of $\gamma$ is critical in the sense that half of the discoveries are lost when moving from $\gamma=0.05$ to $\gamma=0.01$. This is not surprising, and has been already seen in the simulation study. On the other hand, the impact of $\alpha$ in the decision making is less important; this is partly explained by the large number of tests, which results in a small variance term. The results provided by the several NSE-controlling methods are similar, with the only exception of the conservative method $N^\alpha(\gamma;1)$, which always results in a visibly smaller amount of rejected nulls. All of this agrees with the findings in the simulations.\\

Just to discuss the results of one of the methods, consider for instance $N_S^\alpha(\gamma)$ with $\gamma=\alpha=0.05$. This method indicates that, 95\% confident, there are at least 480 effects among the 606 tests with p-value below $\gamma=0.05$, and proceeds to reject the 480 nulls with the smallest p-values. The probability that the number of false positives exceeds the number of false negatives among the significant p-values is bounded by $0.05$; in particular, the number of false positives is smaller than $606-480=126$ with confidence above 95\%. The application of Benjamini-Hochberg (BH) FDR-controlling procedure at level $\alpha=0.05$ resulted in 94 rejected nulls, a number that was increased to only 159 or 157 when using the adaptive BH based on $\hat \pi_0^S$ or $\hat \pi_0^D$, respectively. This illustrates how the number of findings may be enlarged through the application of NSE-controlling methods, compared to the approach which pre-specifies a bound for the expected proportion of false discoveries.\\

In practice, it is of interest to evaluate both the expected proportion of false discoveries and the proportion of non-true nulls which are rejected (i.e. the statistical power). These quantities can be estimated by

\begin{equation*}
	\mbox{FDR}_m(x)=\frac{m \hat \pi_0 x}{N}
\end{equation*}

\noindent and

\begin{equation*}
	\mbox{pow}_m(x)=\frac{N-m\hat \pi_0 x}{m(1-\hat \pi_0)}
\end{equation*}

\noindent where $x$ is the threhold p-value, $N$ denotes the amount of rejected nulls (i.e. the number of p-values below $x$), and $\hat \pi_0$ is an estimator for the proportion of the true nulls. For instance, for the aforementioned decision of $N_S^\alpha(\gamma)$ with $\gamma=\alpha=0.05$, we obtain $\mbox{FDR}_m(0.0324)=0.1448$ and $\mbox{pow}_m(0.0324)=0.4001$, where $\hat \pi_0^S$ was used in the place of $\hat \pi_0$. These figures reduce to $0.0720$ and $0.2126$ when choosing $\gamma=0.01$ and $\alpha=0.05$. Accordingly, the (estimated) expected number of false discoveries for $N_S^\alpha(\gamma)$ is 69 out of 480 ($\gamma=0.05$), or 17 out of 234 ($\gamma=0.01$).

	\begin{table}
		\begin{center}
				\caption{Number of rejected nulls (Rej) and threshold p-value (pval) for NSE controlling procedures with admissibility significance threshold $\gamma$ and level $\alpha$. Hedenfalk data.}
		\begin{tabular}{ ccccccccc } 
			\hline
			$\gamma=$&	$0.05$ & & & & $0.01$ & & &\\
			\hline
			$\alpha=$	& $0.05$ & & $0.01$& & $0.05$ & &$0.01$ &\\
			\hline
			Method & 	Rej & pval & Rej & pval & Rej & pval & Rej & pval \\
			\hline
			$N^\alpha(\gamma;1)$& 427	 & 0.0254  & 418  & 0.0243 & 224 & 0.0072 & 220 & 0.0070\\ 
			$N^\alpha(\gamma,\hat \pi_0^S)$& 482	 &0.0326  & 475  & 0.0316 & 235 & 0.0079 & 232 & 0.0076\\ 
			$N^\alpha(\gamma;\hat \pi_0^D)$	& 475	 & 0.0316  & 467  & 0.0303 & 234 & 0.0078 & 231 & 0.0075\\ 
			$N_S^\alpha(\gamma)$	& 480	 & 0.0324  & 473  & 0.0315 & 235 & 0.0079 & 232 & 0.0076\\ 
			$N_D^\alpha(\gamma)$	& 473	 & 0.0315  & 466  & 0.0303 & 234 & 0.0078 & 231 & 0.0075\\ 
			\hline
		\end{tabular}
	\end{center}
		\label{tab:hedenfalk}
	\end{table}

\begin{figure}[!h]
	\centering			
	\includegraphics[width=80mm]{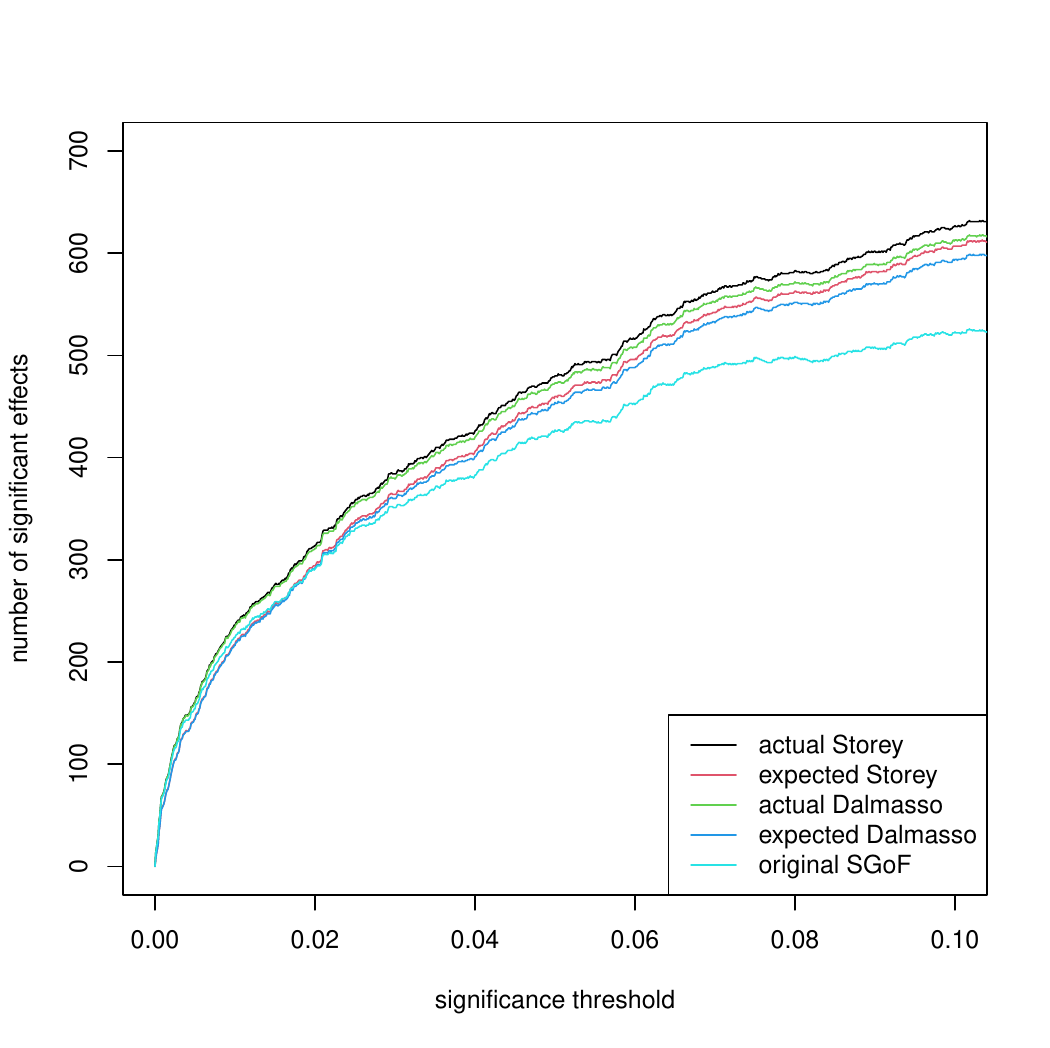}
	\caption{Number of significant effects declared by each method at level $\alpha=0.05$ along a grid of $\gamma$-significance thresholds. Hedenfalk data.}
	\label{fig1}
\end{figure}

Figure \ref{fig1} depicts de decision, level $\alpha=0.05$, on the number of $\gamma$-significant effects obtained from several methods, along a grid of $\gamma$-values on the interval $(0,0.1)$. Methods $N_S^\alpha(\gamma)$ and $N_D^\alpha(\gamma)$ (black and green lines) are in well agreement, particularly for small $\gamma$. Original SGoF method, which as mentioned coincides with $N^\alpha(\gamma;1)$, exhibits a more conservative behavior. The results of two other methods are displayed for comparison purposes. The red line corresponds to a modification of $N_S^\alpha(\gamma)$ that focuses on the expected, rather than on the actual, number of $\gamma$-significant effects, following the spirit of SGoF$_1$ procedure in de U\~na-\'Alvarez (2011); see Remark 3 in the Appendix for details. The blue line corresponds to an analogous modification for $N_D^\alpha(\gamma)$, see Remark 4 in the Appendix. For Hedenfalk data these latter methods report a smaller number of discoveries, which is in agreement with the theoretical discussion in the Appendix. Indeed, their power is even dominated by that of SGoF when $\gamma<0.02$.


\subsection{Diz data}

As a second illustrative example, we consider the relatively small set of $m=261$ p-values coming from the experiment in eggs of \textit{Mytilus edulis} (a marine mussel) performed by Diz et al. (2009). In the experiment, the expression profiles of two different lines of \textit{M. edulis} were compared. In this case, the estimated proportion of true nulls was 0.8427 ($\hat \pi_0^D$) or 0.7969 ($\hat \pi_0^S$). There were $26$ p-values below the significance threshold $\gamma=0.05$. The number of rejections of NSE-controlling methods for $\gamma=0.05$ at level $\alpha=0.05$ ranged between 7 and 10, where the most conservative decision corresponded to $N^\alpha(\gamma;1)$. The two refined methods $N_S^\alpha(\gamma)$ and $N_D^\alpha(\gamma)$ reported 10 and 9 rejections respectively. The estimated FDR for $N_S^\alpha(\gamma)$, based on the estimator $\hat \pi_0^S$, was 0.2305 (power: 0.1452), meaning an expected number of false discoveries between 2 and 3. The full set of results for $\gamma, \alpha \in \{0.05,0.01\}$ is reported in Table 7. For this dataset, FDR control at level $\alpha=0.05$ resulted in no discovery, even when using the adaptive BH method based on any of the estimators for $\pi_0$.\\

	\begin{table}
		\begin{center}
					\caption{Number of rejected nulls (Rej) and threshold p-value (pval) for NSE controlling procedures with admissibility significance threshold $\gamma$ and level $\alpha$. Diz data.}
		\begin{tabular}{ ccccccccc } 
			\hline
			$\gamma=$&	$0.05$ & & & & $0.01$ & & &\\
			\hline
			$\alpha=$	& $0.05$ & & $0.01$& & $0.05$ & &$0.01$ &\\
			\hline
			Method & 	Rej & pval & Rej & pval & Rej & pval & Rej & pval \\
			\hline
			$N^\alpha(\gamma;1)$&	7 &  0.0777& 4 &0.0037 & 2&0.0035 &1 &0.0009\\ 
			$N^\alpha(\gamma,\hat \pi_0^S)$&10	 &0.0111  & 8 &0.0093 & 3&0.0037 &2 &0.0035\\ 
			$N^\alpha(\gamma;\hat \pi_0^D)$	&9	 &0.0102  & 7 &0.0077 & 3& 0.0037& 2&0.0035\\ 
			$N_S^\alpha(\gamma)$	&	10 & 0.0111 & 7 & 0.0077& 3& 0.0037& 2&0.0035\\ 
			$N_D^\alpha(\gamma)$	&	9 &  0.0102&6  & 0.0068& 3& 0.0037& 2&0.0035\\ 
			\hline
		\end{tabular}
		\label{tab:diz}
	\end{center}
	\end{table}

\begin{figure}[htb]
	\begin{center}			
	\includegraphics[width=80mm]{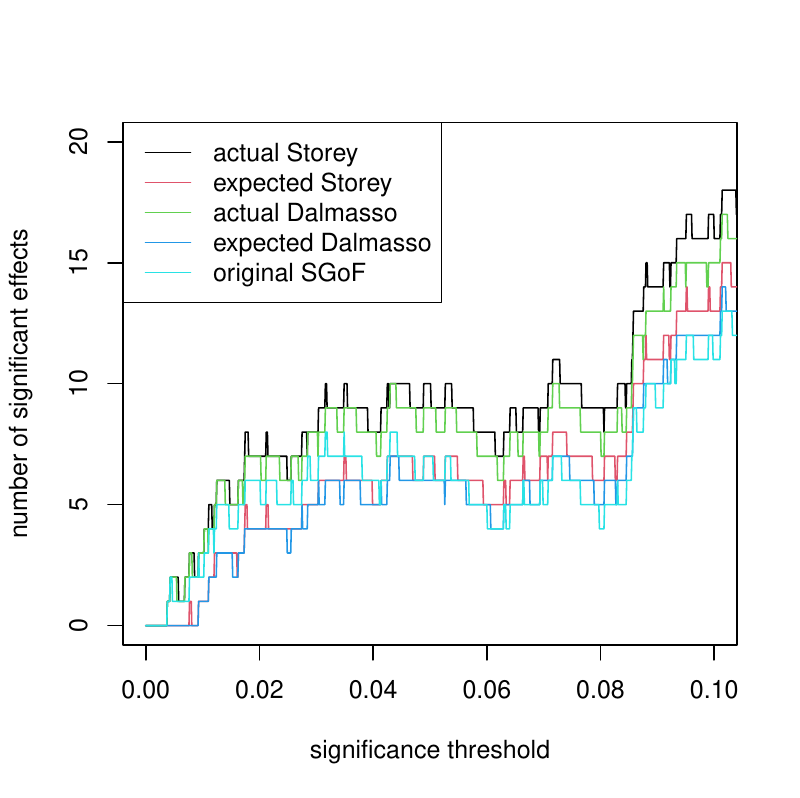}
	\caption{Number of significant effects declared by each method at level $\alpha=0.05$ along a grid of $\gamma$-significance thresholds. Diz data.}
	\end{center}
	\label{fig2}
\end{figure}

Figure \ref{fig2} reports the number of rejected nulls at level $\alpha=0.05$ for each method; a grid of $\gamma$-values on the interval $(0,0.1)$ was used. As for Hedenfalk data, see Figure \ref{fig1}, the number of rejections for Diz data is maximum for Storey-type data-driven and Dalmasso-type data driven methods, which dominate original SGoF procedure. The methods that focus on the expected amount of significant effects are less powerful than those targeting the actual amount.

\section{Discussion}

Multiple testing methods that bound the FWER or the FDR by a given level $\alpha$ may be unable to identify a single effect. This may be due to the presence of a small amount of non-true nulls, or because the true alternatives are close to their corresponding nulls. On the oher hand, the researcher may hesitate about which $\alpha$ to use. In such circumstances, alternative error criteria are welcome. NSE-controlling procedures report a number of rejected nulls that, with probability $1-\alpha$, will not be too large. They achieve that goal by, first, focusing on the nulls with p-values below a certain significance threshold $\gamma$ and, second, reporting a $(1-\alpha)$-lower bound for the number of effects among the referred subset. The role of $\gamma$ is to control for the type I error for each particular test; on its turn, $\alpha$ bounds the probability for the false positives to exceed the number of false negatives among the significant p-values. Compared to FWER- or FDR-based methods, statistical power may be enhanced when using NSE control, which does not fix any preliminary overall bound for the type I error.\\

Interestingly, we have seen how NSE-controlling methods can be computed merely from the expected proportion of true nulls ($\pi_0$) and the actual proportion of p-values below the individual significance threshold ($F_m(\gamma)$). In practice, $\pi_0$ can be estimated from the p-values themselves; NSE control is still guaranteed by updating the corresponding variance term. The simplicity of the proposed methods is one of their main appealings, together with their ability to inform on the presence of real effects in scenarios where other error criteria may fail because they are too strict.\\

Another property of NSE control is that it focuses on the actual number of effects rather than on the expected amount. Therefore, the results are adaptive, in the sense that they are valid for the problem at hand, without taking care of population means which may no be representative of the actual sample. This makes a difference with respect to other existing methods that provide a $(1-\alpha)$-lower bound for the expected proportion of non-true nulls with p-value below $\gamma$, like for instance SGoF$_1$ method (de U\~na-\'Alvarez, 2011). Indeed, under NSE control the number of rejections may be larger than the referred expected amount with probability greater than $\alpha$. On the other hand, original SGoF pivots on the complete null, while plug-in NSE-controlling methods estimate the expected number of significant p-values under any configuration of the null hypotheses.
These remarks help to understand why NSE control may enhance the power compared to other existing procedures.\\

The proposed methods can be refined through the application of improved estimators for $\pi_0$. Each employed estimator requires however the investigation of the resulting variance term. Further reductions of the estimation bias for the proportion of true nulls would be of great interest in order to fit the nominal coverage $1-\alpha$ more accurately. The usual bias-variance trade-off must be considered in any case to ensure power gains. Extensions of the proposed NSE-controlling methods include corrections for possibly dependent or discrete p-values. These topics are left for our future research.

\begin{acknowledgement}
Work supported by the grant PID2020-118101GB-I00, Ministerio de Ciencia e Innovaci\'on (MCIN/ AEI /10.13039/501100011033).
\end{acknowledgement}
\vspace*{1pc}

\noindent {\bf{Conflict of Interest}}

\noindent {\it{The authors have declared no conflict of interest.}}

\section*{Appendix}



In this section the proofs to the several results in the paper are provided.\\

\textbf{Proof to Proposition 2.} Introduce

$$T_m=F_m^0(\gamma)-\gamma \hat \pi_0^S=\frac{1}{m}\sum_{i=1}^m \left( I(p_i\leq \gamma,H_{0i}=0)-\frac{\gamma}{1-\lambda}I(p_i>\lambda)\right) \equiv \frac{1}{m}\sum_{i=1}^m \xi_i.$$

\noindent It holds $E(T_m)=E(\xi_1)$ and $V(T_m)=m^{-1}V(\xi_1)=m^{-1}(E(\xi_1^2)-E^2(\xi_1))$. Now, by recalling $F^0(\gamma)\equiv P(p_1\leq \gamma,H_{01}=0)=\gamma \pi_0$, one gets

$$E(\xi_1)=\gamma \pi_0-\frac{\gamma}{1-\lambda}(1-F(\gamma)) \equiv \mu_T$$

\noindent and

$$E(\xi_1^2)=\gamma \pi_0+\frac{\gamma^2}{(1-\lambda)^2}(1-F(\lambda))-\frac{2\gamma}{1-\lambda}I(\lambda<\gamma)(\gamma-\lambda)\pi_0.$$

\noindent Then, with $\sigma_T^2 \equiv E(\xi_1^2)-\mu_T^2=\sigma^2_S(\gamma;\pi_0)$, by the Central Limit Theorem it holds

$$P(m^{1/2}\frac{T_m-\mu_T}{\sigma_T} \leq z_\alpha) \rightarrow 1-\alpha$$

\noindent as $m \rightarrow \infty$. Now, since $F_m^0(\gamma)=F_m(\gamma)-F_m^1(\gamma)$,

$$P(F_m^1(\gamma)\geq F_m(\gamma)-\gamma \hat \pi_0^S - m^{-1/2}\sigma_T z_\alpha)=P(m^{1/2}\frac{T_m-\mu_T}{\sigma_T}\leq z_\alpha-m^{1/2}\frac{\mu_T}{\sigma_T}).$$

\noindent Since $\mu_T=-\gamma(1-\pi_0)(1-F^{(1)}(\lambda))/(1-\lambda)\leq 0$, probability above is lowerly bounded by $P(m^{1/2}(T_m-\mu_T)/\sigma_T \leq z_\alpha)$. This ensures the asymptotic NSE control stated by Proposition 2. Indeed, theoretically the probability approaches 1 as $m$ grows; however, $\mu_T$ is almost vanishing in practical settings (and, in particular, it is exactly zero when $\pi_0=1$), so one may need a huge number of hypotheses to see departures from the nominal level $1-\alpha$.\\

\textbf{Remark 1.} The theoretical values of $-\mu_T$ ($\times10^3$) and $\sigma_T$ in the simulated settings are reported in Table \ref{tab:musigma}. It is seen that $\mu_T$ is small, and that it decreases as $\mu$ and $\pi_0$ increase (scenario $\pi_0=1$ is skipped since $\mu_T=0$ in that case).

	\begin{table}
		\begin{center}
				\caption{Values of $-10^3\mu_T$ and $\sigma_T$ (in brackets) for the simulated models. Cases $\gamma=0.05$ and $\gamma=0.01$. The conditional cdf $F^{(1)}$ was approximated from a single sample with $10^6$ p-values.}
		\begin{tabular}{ cccccc } 
			\hline
			& &	Storey & & Dalmasso & \\
			\hline
			& 	& $\pi_0=0.9$ & $\pi_0=0.8$& $\pi_0=0.9$& $\pi_0=0.8$\\
			\hline
			$\gamma=0.05$ &$\mu=1$&	$0.57243$  & $1.1449$ & $0.81191$ & $1.6238$\\ 
			& &	$(0.22261)$ &$(0.21003)$  & $(0.22234)$ & $(0.20975)$\\ 
			& $\mu=1.5$	&	$0.03596$ & $0.07192$ &  $0.10616$&$0.21232$\\ 
			& &	$(0.22249)$ & $(0.20978)$ & $(0.22224)$ &$(0.20953)$\\ 
			&	$\mu=2$	&	$0.00066$ & $0.00132$ & $0.008190$ &$0.01638$ \\ 
			& 	&$(0.22249)$	 & $(0.20976)$ & $(0.22223)$ & $(0.20952)$\\
			\hline
			$\gamma=0.01$& $\mu=1$&	$0.11449$  & $0.22897$ &  $0.16238$& $0.32477$\\ 
			& &$(0.09582)$	 &$(0.09036)$  & $(0.09582)$ & $(0.09035)$\\ 
			& $\mu=1.5$	&$0.00719$	 & $0.01438$ & $0.02123$ &$0.04246$\\ 
			& &	$(0.09581)$ & $(0.09033)$ & $(0.09581)$ &$(0.09033)$\\ 
			& $\mu=2$	&$0.00013$	 & $0.00026$ & $0.00164$ & $0.00328$\\ 
			& &$(0.09581)$	 & $(0.09033)$ & $(0.09581)$ & $(0.09033)$\\
			\hline
		\end{tabular}
		\label{tab:musigma}	
	\end{center}
	\end{table}

\textbf{Remark 2.} The conservativeness of SGoF procedure compared to NSE control with data-driven $\pi_0$ comes from equality

\begin{eqnarray*}
	P\left(F_m^1(\gamma)\geq F_m(\gamma)-\gamma-z_\alpha(\gamma(1-\gamma)/m)^{1/2}\right)=\\
	P\left(m^{1/2}\frac{F_m^0(\gamma)-F^0(\gamma)}{\gamma \pi_0(1-\gamma \pi_0)}\leq z_\alpha c_1 +m^{1/2}c_2\right)
\end{eqnarray*}

\noindent where

$$c_1=\left[ \frac{\gamma(1-\gamma)}{\gamma \pi_0(1-\gamma \pi_0)}\right]^{1/2}\hspace{0.2 cm}\mbox{and}\hspace{0.2 cm}c_2=\frac{\gamma(1-\pi_0)}{(\gamma \pi_0(1-\gamma \pi_0))^{1/2}}.$$

\noindent The term $m^{1/2}c_2$, which comes from the potential bias attached to SGoF choice $\pi_0=1$, leads to a probability visibly larger than the nominal $1-\alpha$ whenever $\pi_0<1$. The constant $c_1$, which is larger than 1 when $\gamma<1/2$ and which comes from a wrong calibration of the variance, contributes to this departure too, although its impact is relatively small.\\

\textbf{Proof to Proposition 3.} Similarly as in the proof to Proposition 2, introduce

$$T_m=F_m^0(\gamma)-\gamma \hat \pi_0^D=\frac{1}{m}\sum_{i=1}^m \left( I(p_i\leq \gamma,H_{0i}=0)+\gamma \log(1-p_i)\right) \equiv \frac{1}{m}\sum_{i=1}^m \xi_i.$$

\noindent As in the proof to Proposition 2, it holds $E(T_m)=E(\xi_1)$ and $V(T_m)=m^{-1}V(\xi_1)=m^{-1}(E(\xi_1^2)-E^2(\xi_1))$, where here

$$E(\xi_1)=\gamma \pi_0+\gamma \int \log(1-x)dF(x) \equiv \mu_T$$

\noindent and

$$E(\xi_1^2)=\gamma \pi_0+\gamma^2\int \log^2(1-x)dF(x)+2\gamma \pi_0\int_0^\gamma\log(1-x)dx.$$

\noindent Since $\int_0^\gamma \log(1-x)dx=-(1-\gamma)\log(1-\gamma)-\gamma$, we get $\sigma_T^2 \equiv E(\xi_1^2)-\mu_T^2=\sigma^2_D(\gamma;\pi_0)$. By the Central Limit Theorem it holds

$$P(m^{1/2}\frac{T_m-\mu_T}{\sigma_T} \leq z_\alpha) \rightarrow 1-\alpha$$

\noindent as $m \rightarrow \infty$. Similarly as for Proposition 2, we get

$$P(F_m^1(\gamma)\geq F_m(\gamma)-\gamma \hat \pi_0^D - m^{-1/2}\sigma_T z_\alpha)=P(m^{1/2}\frac{T_m-\mu_T}{\sigma_T}\leq z_\alpha-m^{1/2}\frac{\mu_T}{\sigma_T})$$

\noindent where here $\mu_T=\gamma(1-\pi_0)\int \log(1-x)dF^{(1)}(x)\leq 0$. Then, probability above is lowerly bounded by $P(m^{1/2}(T_m-\mu_T)/\sigma_T \leq z_\alpha)$, and NSE control is asymptotically ensured. Again, $\mu_T=0$ when $\pi_0=1$ and, hence, the asymptotic NSE confidence level of the method is exactly $1-\alpha$ in this case. This confidence level can be larger than the nominal $1-\alpha$ in settings with $\pi_0<1$ although, since $\mu_T$ is often almost vanishing, only small departures are expected.\\

\textbf{Remark 3.} NSE control does not provide a $(1-\alpha)$-confidence lower bound for the \textit{expected} number of $\gamma$-significant cases, $F^1(\gamma)$. This would have required a larger variance term. In order to illustrate this, consider for instance the Storey-based method $N_S^\alpha(\gamma)$. It holds

\begin{eqnarray*}
	P\left(F^1(\gamma) \geq N_S^\alpha(\gamma)\right)=P\left( m^{1/2}\frac{F_m(\gamma)-\gamma \hat \pi_0^S-F^1(\gamma)}{\sigma_S(\gamma;\pi_0)}\leq z_\alpha\right).
\end{eqnarray*}

\noindent While the expected value of $F_m(\gamma)-\gamma \hat \pi_0^S-F^1(\gamma)$ is negligible (it is just the $\mu_T$ parameter in the proof to Proposition 2), the variance is not well calibrated. Actually, $m$ times the variance of $F_m(\gamma)-\gamma \hat \pi_0^S$ is

$$F(\gamma)+\frac{\gamma^2}{(1-\lambda)^2}(1-F(\lambda))-(\mu_T+F^1(\gamma))^2,$$
\noindent provided that (to simplify things) $\gamma\leq \lambda$. By using $\mu_T \approx 0$ and $F(\gamma)-F^1(\gamma)^2\geq F(\gamma)-F^1(\gamma)=\gamma\pi_0$ it is seen that this variance is larger than $\sigma_S^2(\gamma;\pi_0)$. A similar result can be obtained for the Dalmasso-based method $N_D^\alpha(\gamma)$; see Remark 4 below. The variance increase explains the results provided in our simulation study, and in the real data illustrations too. All of this suggests a modification of original SGoF procedure to control for the expected number of significant cases. Indeed, de U\~na-\'Alvarez (2011) already discussed the idea of replacing $F_m(\gamma)-\gamma$ in SGoF by $F_m(\gamma)-\gamma \hat \pi_0$ for some estimate $\hat \pi_0$, and coined the term SGoF$_1$ for the modified method. Still, as mentioned, focusing on the expected (rather than on the actual) number of significant effects reduces the power; NSE-control is recommended.\\

\textbf{Remark 4.} For Dalmasso-based criterion $F_m(\gamma)-\gamma \hat \pi_0^D$ the variance is $1/m$ times

$$F(\gamma)+\gamma^2\int \log^2(1-x)dF(x)+2\gamma\int_0^\gamma \log(1-x)dF(x)-(\mu_T+F^1(\gamma))^2.$$

\noindent Use $\mu_T \approx 0$, $F(\gamma)-F^1(\gamma)^2 \geq \gamma \pi_0$, $dF(x)=\pi_0dx+(1-\pi_0)dF^{(1)}(x)$ in the second integral, and $\gamma(1-\pi_0)\int_0^\gamma \log(1-x)dF^{(1)}(x)\geq \mu_T$ to lowerly bound the variance above by $\sigma^2_D(\gamma;\pi_0)$. This shows that more power can be reached when focusing on the actual (rather than on the expected) number of significant effects.

\end{document}